\documentclass{icrc}
\usepackage{graphicx}

\def\grays{$\gamma$-rays}

\begin{document}

\title{New calculation of radioactive secondaries in cosmic rays}
\author[1,2]{I. V. Moskalenko}
\affil[1]{NRC-NASA/Goddard Space Flight Center, Code 660, Greenbelt, MD 20771, U.S.A.}
\affil[2]{Institute of Nuclear
   Physics, M.\ V.\ Lomonosov Moscow State University, 119 899 Moscow, Russia}
\author[3]{S. G. Mashnik}
\affil[3]{Theoretical Division, Los Alamos National Laboratory, Los Alamos, NM 97545, U.S.A.}
\author[4]{A. W. Strong}
\affil[4]{Max-Planck-Institut f\"ur extraterrestrische Physik, 
Postfach 1312, 85741 Garching, Germany}

\correspondence{imos@milkyway.gsfc.nasa.gov}

\firstpage{1}
\pubyear{2001}

\maketitle

\begin{abstract}
We use a new version of our numerical model for particle propagation in
the Galaxy to study radioactive secondaries.
For evaluation of the production
cross sections we use the Los Alamos compilation of all available
experimental cross sections together with calculations using 
the improved Cascade-Exciton Model code CEM2k. Using the radioactive
secondary ratios $^{26}$Al/$^{27}$Al, $^{36}$Cl/Cl, $^{54}$Mn/Mn, 
we show how the improved cross-section calculations
together with the new propagation code allow us to better 
constrain the size of the CR halo.
\end{abstract}

\section{Introduction}
%#########################################################################

In recent years, new and accurate data have become available in  CR
astrophysics; more CR experiments are planned for launch in several
years  that will tremendously increase the quality and accuracy of CR
data making further progress dependent on detailed models. Data
will continue to flow from the high resolution detectors on Ulysses,
Advanced Composition Explorer (ACE) and Voyager space missions.
Several high resolution space experiments will be in orbit in the
nearest 2--3 years, e.g., PAMELA to measure antiprotons, positrons,
electrons, and isotopes H through C over the energy range of 0.1 to
200 GeV, and Alpha Magnetic Spectrometer (AMS) to measure particle and
nuclear spectra to TeV enegies.

Measurements of secondary stable and radioactive nuclei in CR provide
basic information necessary to probe  large-scale Galactic properties,
such as the diffusion coefficient and halo size, as well as mechanisms
and sites of CR acceleration. Meanwhile, the accuracy of many of the nuclear
cross sections used in CR astrophysics is far behind the
accuracy of CR measurements of the current missions, such as Ulysses,
ACE, and Voyager, and clearly becomes a factor restricting further
progress.  The widely used semi-phenomenological systematics have
typical uncertainties more than $\sim50$\%, and can sometimes be wrong
by an order of magnitude (Mashnik 2000 and references therein);  this
is reflected in the value of propagation parameters and leads to
uncertainties in the interpretation.

We have previously described a numerical model for the \linebreak[4]
Galaxy encompassing primary and secondary CR, \grays\ and synchrotron
radiation in a common framework \citep{SMR00}.
Up to recently our GALPROP code handled 2 spatial
dimensions, $(R,z)$, together with particle momentum $p$. This was
used as the basis for studies of CR reacceleration, the size of the
halo, positrons, antiprotons, dark matter and the interpretation of
diffuse continuum \grays.

The experience gained from the original version allowed us to design a
new  version of the model, entirely rewritten in C++,  
which incorporates essential improvements over the
older model, and  in which a 3-dimensional spatial grid can be
employed.  It is now  possible to solve the full nuclear
reaction network on the spatially resolved  grid.  We keep however a
``2D'' option since this is often a sufficient approximation and is
much faster to compute than the full 3D case.  The code can thus serve
as a complete substitute for the conventional ``leaky-box'' or
``weighted-slab'' propagation models usually employed, with many
associated advantages such as the correct treatment of radioactive
nuclei, realistic gas and source distributions etc.

In this paper we show our preliminary calculations of the radioactive
secondary ratios $^{26}$Al/$^{27}$Al, $^{36}$Cl/Cl, $^{54}$Mn/Mn
using the Los Alamos compilation of experimental cross sections
together with calculations by the code CEM2k (recognized by the
nuclear physics community as among the best in predictive power as
compared with other similar available codes).

\section{Model}
%#########################################################################

The GALPROP models have been described in full detail elsewhere
\citep{SM98}. The results obtained with the new version of
GALPROP have been discussed in a recent review \citep{SM01a},
and the most recent updates are described in \cite{SM01b}.

In the new version, apart from the option of a full 3D treatment, we
have updated the cross-section code to include latest measurements and
energy dependent fitting functions. The nuclear reaction network is
built using the Nuclear Data Sheets. The beryllium and boron production
was calculated using the authors' fits to major
production cross sections \linebreak C,N,O$(p,x)$Be,B. For the main
channels of production of isotopes of Al, Cl, Mn we use all 
experimental data available to us from the T16 LANL compilation by
\citet{t16lib} and calculations using  the improved version \citep{cem2k}
of the Cascade-Exciton Model \citep{cem} code CEM2k renormalized to
the data if necessary. This code employs sophisticated microphysics via
Monte Carlo calculations and it is difficult to use it
``on-line'' with our propagation code; other cross sections are thus
calculated using the \citet{webber} (wnewtr.for of 1993) phenomenological
approximation renormalized to the data where it exists.
For this purpose we use our internal database consisting of more than 2000
points collected from sources published in 1969--1999. 
(Another option is to use code yieldx\_011000.for by Silberberg and Tsao.)
For calculation of the total inelastic cross
sections we use the latest version of the code CROSEC \citep{crosec}.

The reaction network is solved starting at the heaviest nuclei (i.e.\
$^{64}$Ni), solving the propagation equation, computing all the
resulting secondary source functions, and proceeding to the nuclei
with $A-1$.  The procedure is repeated down to $A=1$.  In this way all
secondary, tertiary etc.\ reactions are  automatically accounted for.
To be completely accurate for all isotopes, e.g.\ for some rare cases
of $\beta^\pm$-decay, the whole loop is repeated twice.

\begin{figure}[!tb]
\includegraphics[width=.45\textwidth]{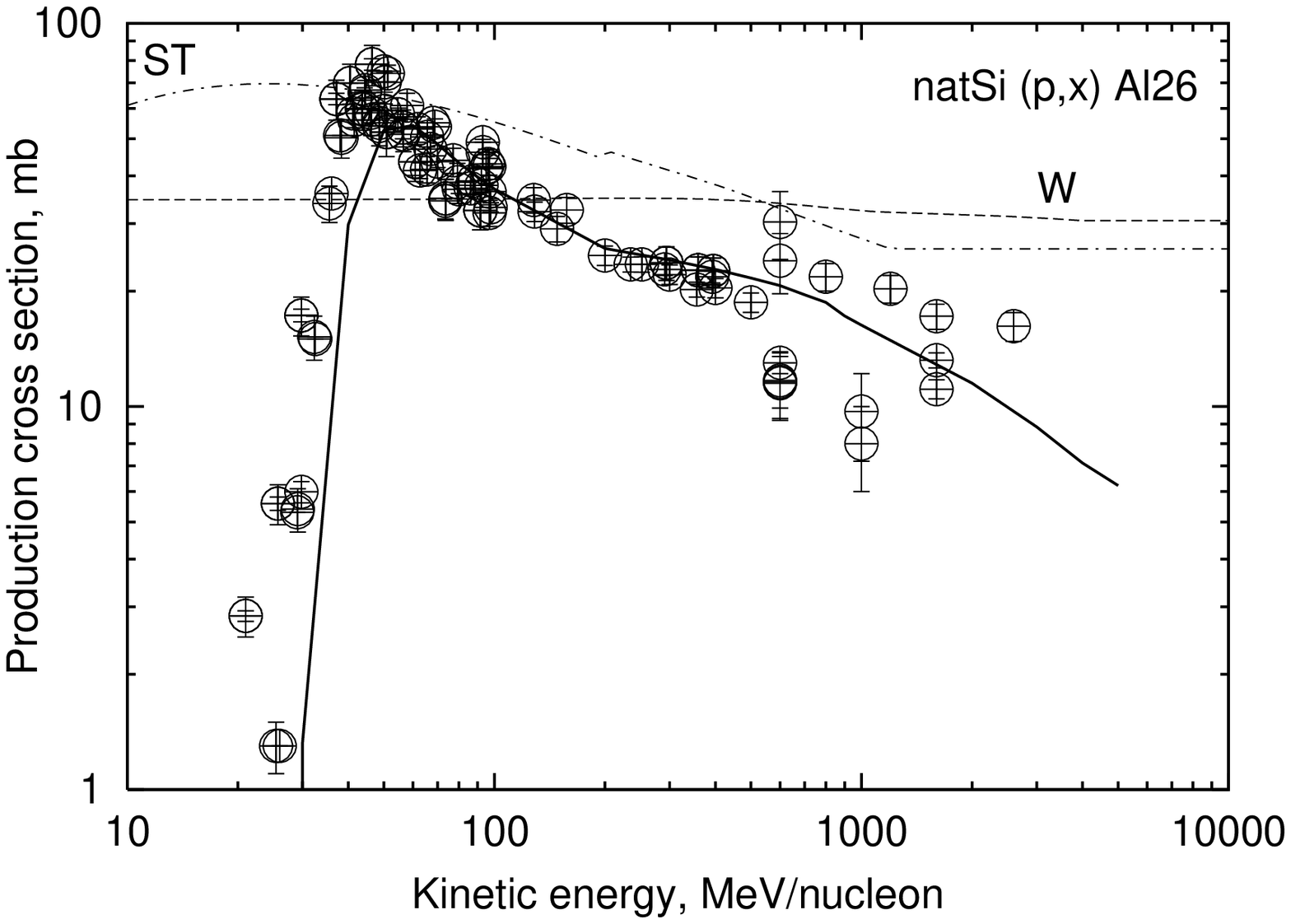}
\includegraphics[width=.45\textwidth]{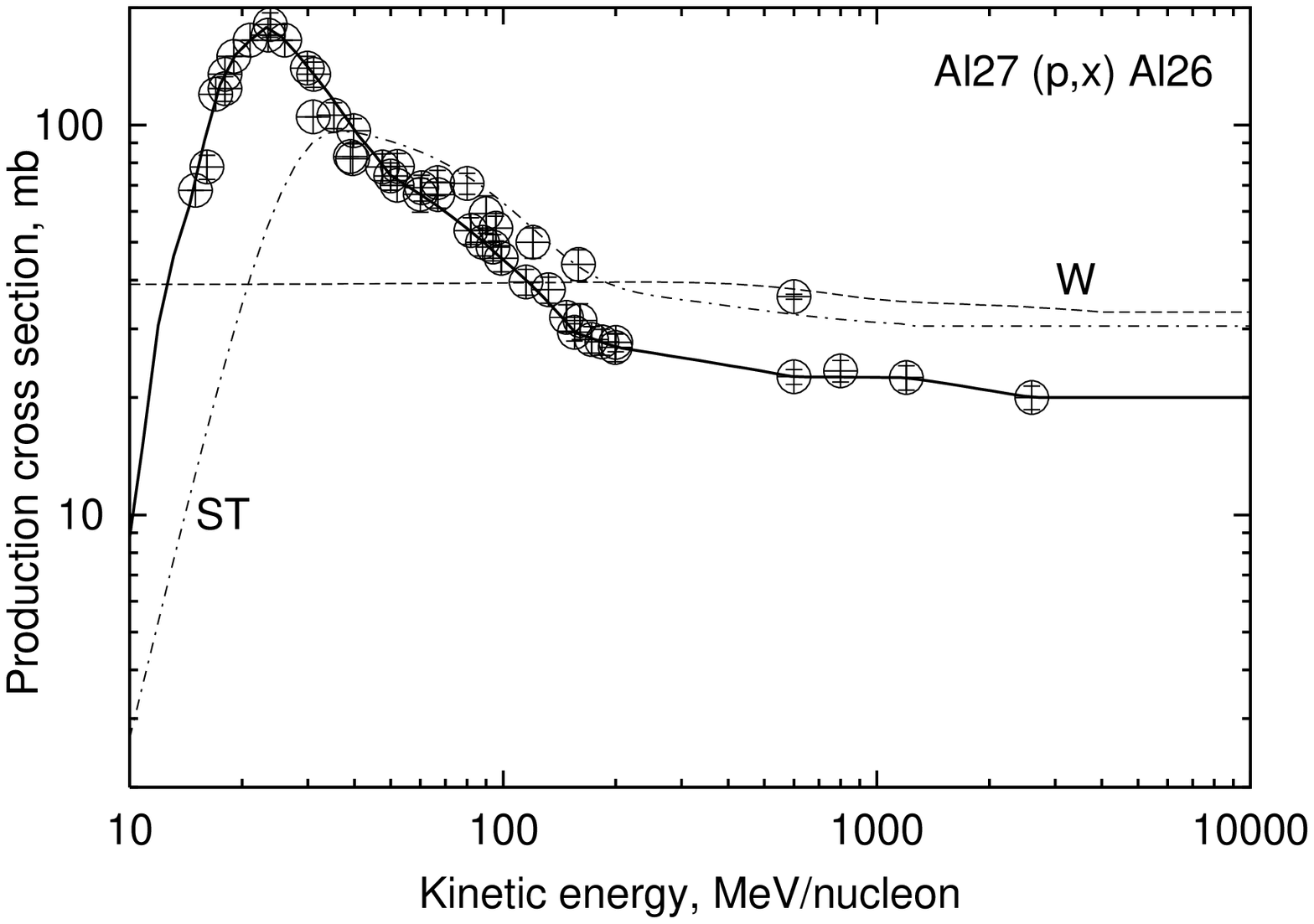}
\includegraphics[width=.45\textwidth]{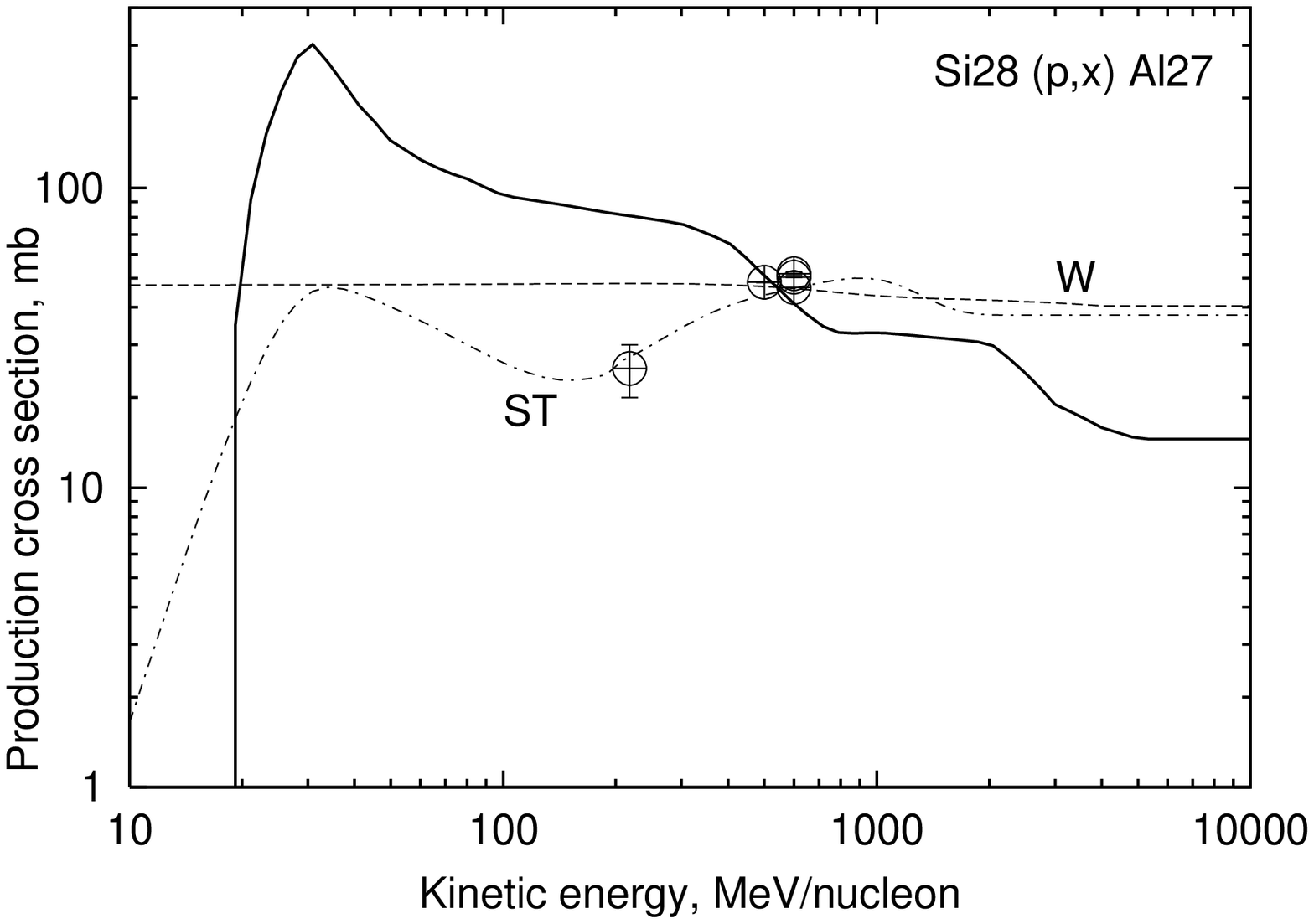}
\caption[fig1a.ps,fig1b.ps,fig1c.ps]{
Production cross sections of Al isotopes.
The line coding: solid line -- our adopted cross section,
dashes -- \citet{webber} code (W), dash-dots -- ST code.
Data: T16 LANL compilation \citep{t16lib}.
} 
\label{fig:cs}
\end{figure}%#################################

\section{Production cross sections}
%#########################################################################

Since the calculations with the modern nuclear codes are very time
consuming we check the effect of the new cross sections only on the
isotopes of Al, Cl, and Mn. The radioactive isotopes of these elements
are the main  astrophysical time clocks which together with stable
secondary isotopes allow us to probe  global Galactic CR properties, 
in particular the  halo size.

For isotopes of these elements we have chosen only the main production
channels  to calculate most accurately.  For $^{26}$Al the main
progenitors are $^{27}$Al and $^{28}$Si, that for $^{27}$Al is
$^{28}$Si. For isotopes of Cl the main progenitor is $^{56}$Fe, but the
contribution of many lighter nuclei is equally important. In the case of
Mn, the main progenitor is $^{56}$Fe with significant contribution of other
isotopes of Fe, except for $^{55}$Mn where only $^{56}$Fe is important.

The experimental data for comparison should be taken carefully.  The
experimental technique in the past ($\gamma$-spectrometry) did not
allow for the individual partial cross section to be extracted.  Those
measured represent the cross section of all the nuclear  reaction
chains ending at the particular isotope, i.e., almost always 
cumulative yields.

The simplest case is production of Al isotopes, however the abundant
experimental measurements exist only for the natural Si $\to ^{26}$Al
reaction. Natural silicon consists of 92\% of $^{28}$Si and the rest
are isotopes $^{29,30}$Si, 5\% and 3\% respectively
\citep{solar_abund}.  The measured cross section of the reaction
$^{nat}$Si$(p,x)^{26}$Al include also
$^{nat}$Si$(p,x)^{26}$Si, but the contribution is small.

Fig.\ \ref{fig:cs} shows the experimental data for the inclusive
reaction $^{nat}$Si$(p,x)^{26}$Al, $^{27}$Al$(p,x)^{26}$Al, and
$^{28}$Si$(p,x)^{27}$Al together with calculations using CEM2k and
\citet{webber} and Silberberg and Tsao (ST) codes. In case of $^{nat}$Si,
calculations include weighted contribution of Si isotopes.  Production
of $^{26}$Al is calculated as the sum of $^{26}$Al and $^{26}$Si
production cross sections.  In this particular case we use the data to
renormalize the $^{28}$Si$(p,x) ^{26}$Al cross sections calculated by
CEM2k.  In case of the reaction $^{28}$Si$(p,x)^{27}$Al ($+ ^{27}$Mg)
we use a fit to the data, though the renormalized CEM2k model also
works well above some tens of MeV.

\section{Propagation of cosmic rays}
%#########################################################################

Our preferred  model for nuclei propagation is that with diffusive reacceleration.
Though it has possible problems with secondary antiprotons and positrons
\citep{moskalenko},\  it describes the spectra of nuclei and the stable  
secondary/pri\-mary nuclei ratios well. 
We thus will use the stochastic reacceleration model (SR-model)
described in \citet{moskalenko}.

As in previous work, for each halo height $z_h$ the model is adjusted to fit 
B/C, and the source abundances at the appropriate energy adjusted to agree 
with the relevant observed stable nuclei ratios;  the 
fluxes of the radioactive isotopes are then computed. 
In this way the uncertainty in the denominator of the ratios is reduced. 
The heliospheric modulation is taken into account using the force-field 
approximation.

\begin{figure}[!t]
\includegraphics[width=.45\textwidth]{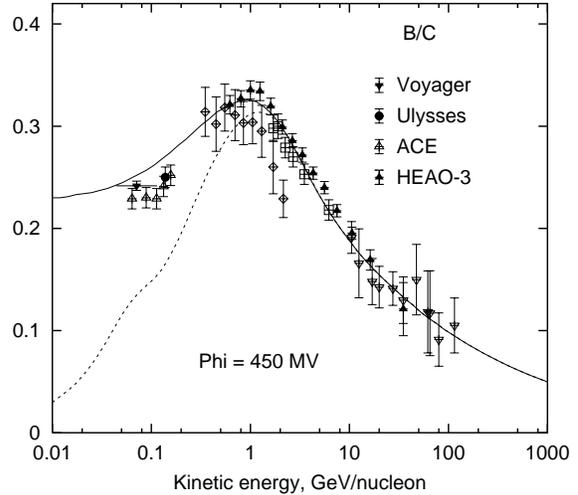}
\caption[fig2.ps]{B/C ratio as calculated for a model with reacceleration.
Upper curve: modulated for 450 MV,
lower curve: interstellar. Data: see \citet{SM01a}.}
\label{fig:bc}
\end{figure}

Fig.\ \ref{fig:bc} shows the predicted interstellar
and modulated B/C ratio compared with observations;
the reacceleration reproduces the peak quite well.

The case of $^{26}$Al/$^{27}$Al was the most uncertain 
giving the largest halo size in \citet{SM01a}. From Fig.\ \ref{fig:cs} it is clear that  
the discrepancy between the cross section calculations and data, which often 
exceeds a factor of 2, introduces a large error in the calculated ratio 
in CR.

\begin{figure}[!t]
\includegraphics[width=.45\textwidth]{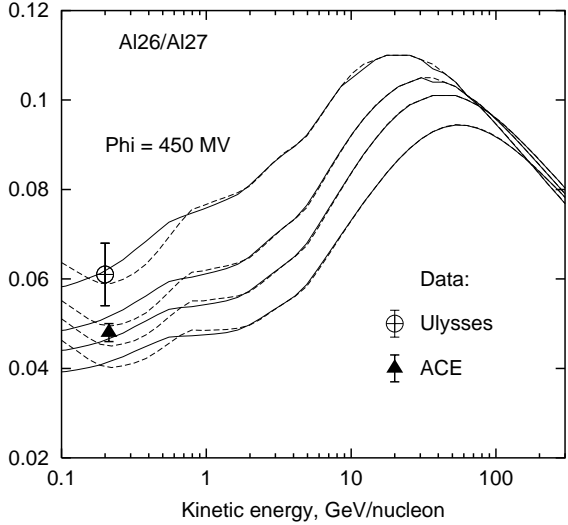}
\caption[fig3.ps]{$^{26}$Al/$^{27}$Al ratio calculated for $z_h=2, 4, 6, 10$ kpc
(top to bottom). Solid curves -- modulated, dashes -- interstellar. Data: ACE -- 
\citet{yanasak}, Ulysses -- \citet{simpson}.}
\label{fig:al}
\end{figure}

\begin{figure}[!t]
\includegraphics[width=.45\textwidth]{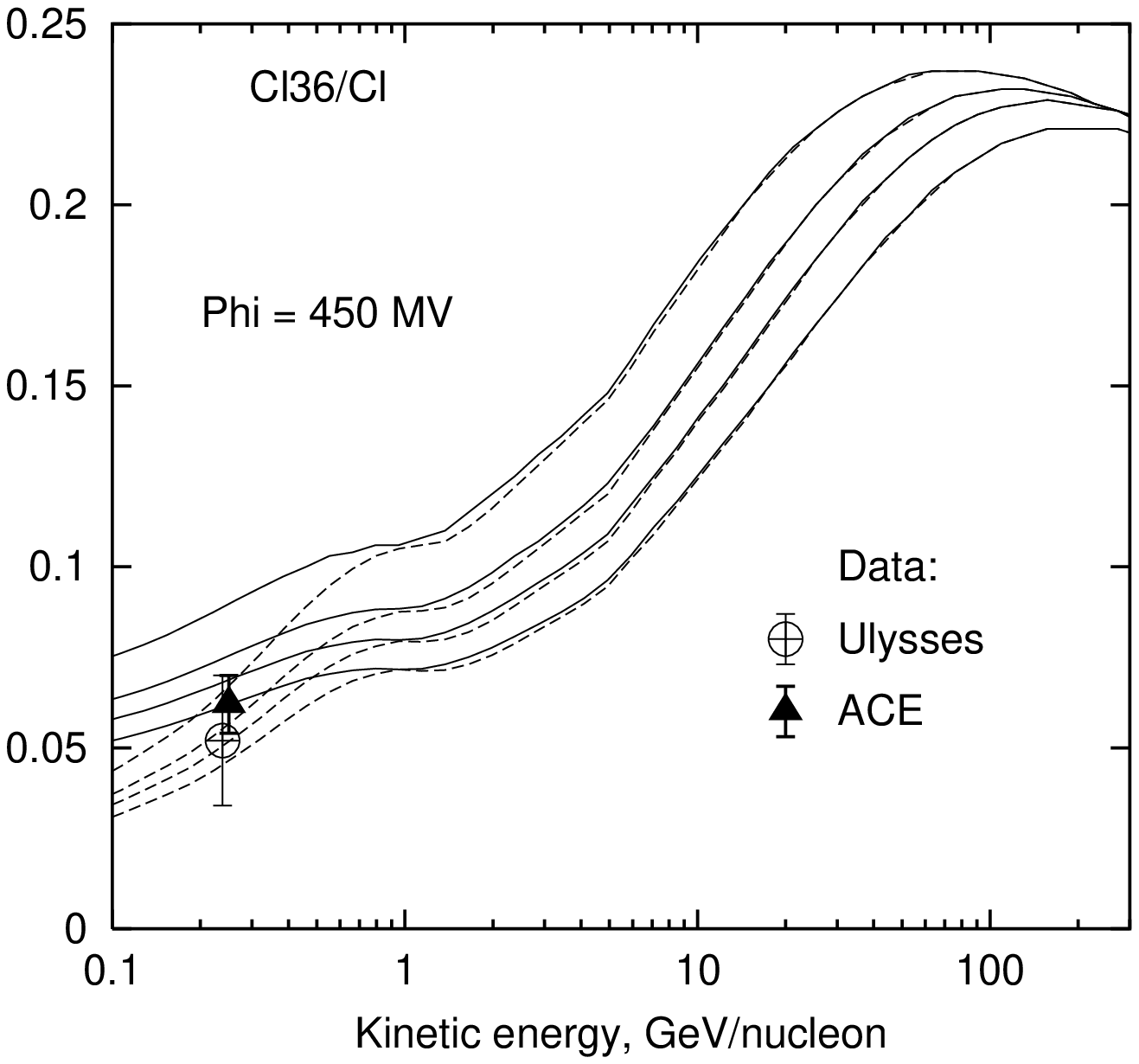}
\caption[fig4.ps]{$^{36}$Cl/Cl ratio calculated for $z_h=2, 4, 6, 10$ kpc
(top to bottom). Solid curves -- modulated, dashes -- interstellar. Data: ACE -- 
\citet{yanasak}, Ulysses -- \citet{connell}.}
\label{fig:cl}
\end{figure}

\begin{figure}[!t]
\includegraphics[width=.45\textwidth]{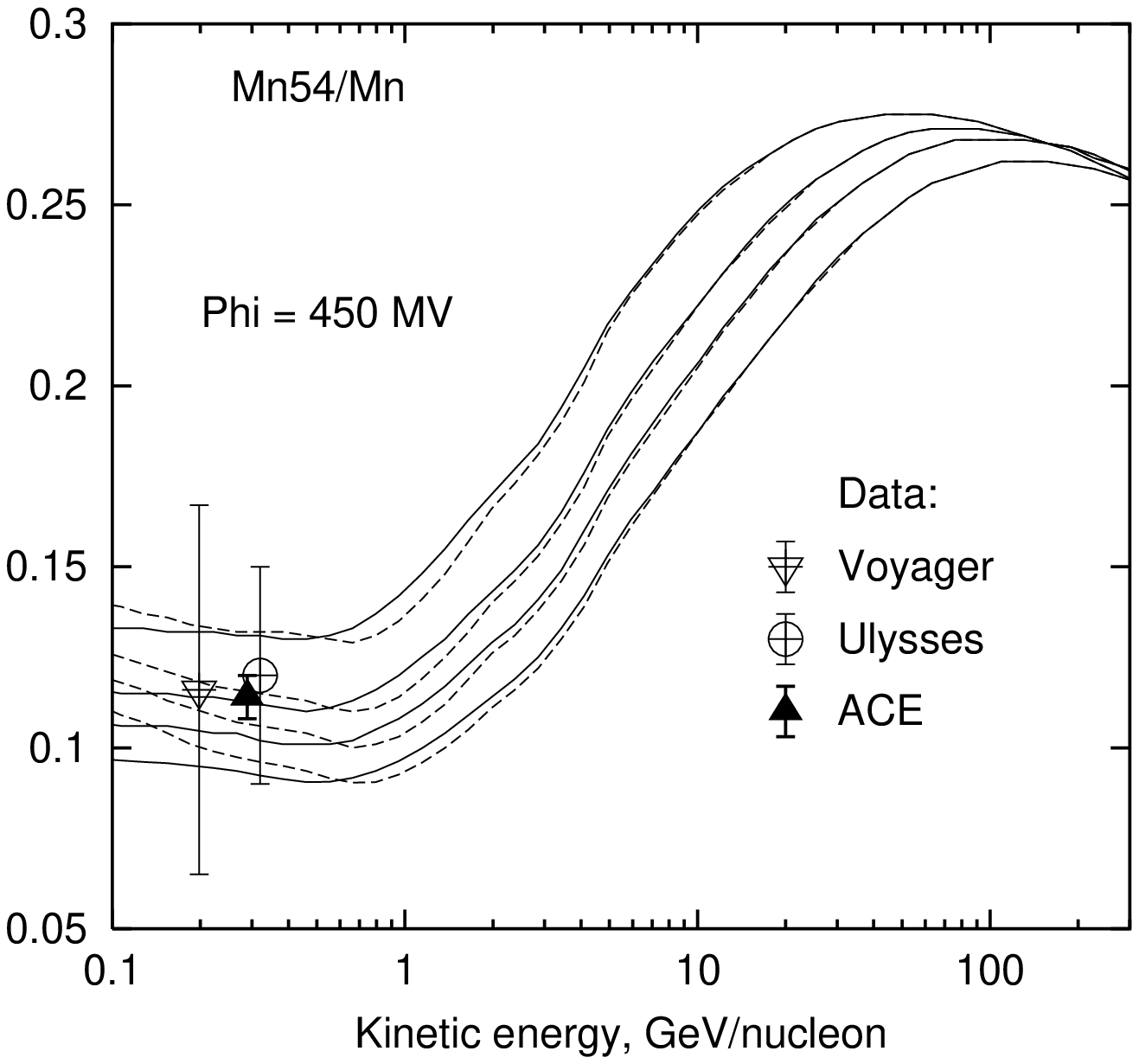}
\caption[fig5.ps]{$^{54}$Mn/Mn ratio calculated for $z_h=2, 4, 6, 10$ kpc
(top to bottom). Solid curves -- modulated, dashes -- interstellar. Data: ACE -- 
\citet{yanasak}, Ulysses -- \citet{duvernois}, Voyager -- 
\citet{lukasiak}.}
\label{fig:mn}
\end{figure}

Figs.\ \ref{fig:al}-\ref{fig:mn} show $^{26}$Al/$^{27}$Al, $^{36}$Cl/Cl,
$^{54}$Mn/Mn ratios calculated with the new cross sections. 
For this calculation we used the half-life times of 0.87 Myr ($^{26}$Al),
0.31 Myr ($^{36}$Cl), 0.63 Myr ($^{54}$Mn).
The ACE data points imply a halo size of a few kpc.
The new limits derived, $^{26}$Al: 3.5--6 kpc, $^{36}$Cl: 4--15 kpc, $^{54}$Mn: 3--7 kpc,
are all consistent with our limits derived from Be: 1.5--6 kpc.
The new limits include the error bars of the elemental abundance
measurements from Ulysses \citep{ulysses}, to which we tune our propagated abundances.
However, for isotopes of Al and Mn they are less important because there is only
one major progenitor in each case. 

\begin{figure}[!t]
\includegraphics[width=.45\textwidth]{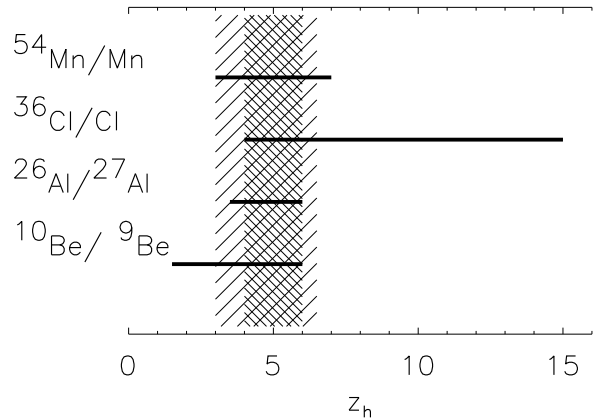}
\caption[fig6.ps]{Halo size limits as derived from the abundances of the four 
radioactive isotopes and ACE data. 
The ranges reflect errors in ratio measurements and source abundances. 
The dark shaded area indicates the range consistent with all ratios;
for comparison the range from  \citep{SM01a} is shown by  light shading.
}
\label{fig:halo}
\end{figure}

Fig.\ \ref{fig:halo} summarizes the halo size constraints 
obtained in this analysis. These estimates are based on the four
radioactive isotopes by requiring consistency
of the calculated ratio with the ACE data \citep{yanasak}
and taking into account the error bars on both prediction and data.
Also shown is the range derived in
\citep{SM01a} employing the \citet{webber}
cross section code.
 
%$^{35,36,37}$Cl/$^{56}$Fe, and $^{53,54,55}$Mn/$^{56}$Fe

To remove uncertainty connected with elemental abundan\-ces errors, we propose
in future to derive ratios of all isotopes of Al, Cl, Mn to the main
progenitor, namely $^{26,27}$Al/$^{28}$Si, $^{35-37}$Cl/$^{56}$Fe, and
$^{53-55}$Mn/$^{56}$Fe, not only the widely used $^{26}$Al/$^{27}$Al, $^{36}$Cl/Cl, and 
$^{54}$Mn/Mn ratios. In case of Al and Mn isotopes this will virtually
eliminate the need to tune the elemental abundances.

Some uncertainty still comes from modulation, 
while the experimental values for the ratios measured by ACE
are rather accurate. However, because of the very flat ratio in the case 
of $^{54}$Mn/Mn (below 1 GeV/nucleon) 
the modulation uncertainty is of minor importance.

The preliminary conclusion to be drawn from
all radioactive nuclei is that, at least within the context
of the present propagation model, $z_h=4-6$ kpc based on the ACE data
and Ulysses elemental abundances.
This is consistent with our previous result $z_h=3-7$ kpc 
\citep{SM01a} and supports our previous conclusion that the large dispersion 
between the isotopes is mostly due to cross-section inaccuracies.

\begin{acknowledgements}
We thank Profs.\ Barashenkov and Polanski for providing us
with their CROSEC code.
This work was partly supported by the NAS/NRC Research 
Associateship Program (I.\ Mos\-kalenko) and the U.S.\ Department of Energy
(S.\ Mashnik).
\end{acknowledgements}

\end{document}